\documentclass[fleqn,twoside]{article}
\usepackage{espcrc2}

\usepackage{graphicx}
\usepackage[figuresright]{rotating}
\usepackage{amsmath}

\def\gsim{\raise0.3ex\hbox{$>$\kern-0.75em\raise-1.1ex\hbox{$\sim$}}}
\def\lsim{\raise0.3ex\hbox{$<$\kern-0.75em\raise-1.1ex\hbox{$\sim$}}}

\title{Cosmic rays and tests of fundamental principles}

\author{Luis Gonzalez-Mestres\address{LAPP, Universit\'e de Savoie, CNRS/IN2P3, B.P. 110, 74941 Annecy-le-Vieux Cedex, France}}

\begin{document}

\begin{abstract}
It is now widely acknowledged that cosmic rays experiments can test possible new physics directly generated at the Planck scale or at some other fundamental scale. By studying particle properties at energies far beyond the reach of any man-made accelerator, they can yield unique checks of basic principles. A well-known example is provided by possible tests of special relativity at the highest cosmic-ray energies. But other essential ingredients of standard theories can in principle be tested: quantum mechanics, uncertainty principle, energy and momentum conservation, effective space-time dimensions, hamiltonian and lagrangian formalisms, postulates of cosmology, vacuum dynamics and particle propagation, quark and gluon confinement, elementariness of particles... Standard particle physics or string-like patterns may have a composite origin able to manifest itself through specific cosmic-ray signatures. Ultra-high energy cosmic rays, but also cosmic rays at lower energies, are probes of both "conventional" and new Physics. Status, prospects, new ideas, and open questions in the field are discussed. The Post Scriptum shows that several basic features of modern cosmology naturally appear in a SU(2) spinorial description of space-time without any need for matter, relativity or standard gravitation. New possible effects related to the spinorial space-time structure can also be foreseen. Similarly, the existence of spin-1/2 particles can be naturally related to physics beyond Planck scale and to a possible pre-Big Bang era.
\vspace{1pc}
\end{abstract}

\maketitle

\section{Introduction}

The formulation, validity domain and experimental tests of fundamental principles of Physics have always been difficult issues requiring long-term work and conceptual evolution. Theoretical ideas and formulations, as well as experimental methods, evolve following this process.

Cosmic-ray experiments are in particular able to detect particles with energies much larger than those that can be produced at man-made accelerators, or having evolved over astrophysical distances and time scales. They therefore play a unique and indispensable role in the exploration and verification of the laws of Physics.  

\subsection{Ether, vacuum and particles}

As early as 1895, Henri Poincar\'e formulated the relativity principle as follows \cite{Poincare1895}: {\it "L'exp\'erience a r\'ev\'el\'e une foule de faits qui peuvent se r\'esumer dans la formule suivante : il est impossible de rendre manifeste le mouvement absolu de la mati\`{e}re, ou mieux le mouvement relatif de la mati\`{e}re par rapport \`{a} l'\'ether. Tout ce qu'on peut mettre en \'evidence, c'est le mouvement de la mati\`{e}re pond\'erable par rapport \`{a} la mati\`{e}re pond\'erable."}

The claimed impossibility to disclose "absolute motion", or even the "relative motion of matter with respect to ether", did not by itself imply considering ether as a real material medium. Poincar\'e explicitly wrote in 1902 \cite{Poincare1902}: {\it "Peu nous importe que l'\'ether existe r\'eellement, c'est l'affaire des m\'etaphysiciens ; l'essentiel pour nous c'est que tout se passe comme s'il existait et que cette hypoth\`{e}se est commode pour l'explication des ph\'enom\`{e}nes. (...) un jour viendra sans doute o\`{u} l'\'ether sera rejet\'e comme inutile."} He therefore considered ether as a practical tool to be possibly abandoned at a later stage of physical theories, but not as a physical entity. Subsequent work by Poincar\'e is to be interpreted basically as the formulation of an effective relativistic geometry of dynamical origin \cite{Gonzalez-Mestres2000a}.

Logunov \cite{Logunov} emphasizes the statement by R.P. Feynman : {\it "It was Poincar\'e's suggestion to make this analysis of what you can do to the equations and leave them alone. It was Poincar\'e's attitude to pay attention to the symmetries of physical laws"}. Today, fundamental symmetries play a central role in standard particle theories up to Planck scale. Furthermore, the evolution of Physics has shown that the interactions of matter can generate new effective symmetries that are exact, for instance, in the low-momentum limit.  

More than a century after the pioneering work by Poincar\'e, Lorentz and other authors, the vacuum of particle physics appears to be a material medium where particle fields condense and whose physical content and structure have direct cosmological implications. The influence of ideas and concepts originating in condensed matter physics has been crucial for this evolution. It has in particular guided the theory of spontaneous symmetry breaking in standard particle theory \cite{Nambu}. 

It has more recently been suggested \cite{gonSL1} that standard relativity may have a composite origin, just as condensed matter can generate low-momentum symmetries of the Lorentz type with the speed of sound playing the role of the critical speed. Disclosing such a composite structure would be possible only at very high energy, most likely through cosmic-ray experiments.   

Between the late 19th century and the early 21st century, the concept of a medium where particles and waves propagate has undergone a deep evolution involving several basic steps. In 1920, having in mind the application of general relativity to macroscopic bodies, Albert Einstein stated about ether \cite{Einstein1920}: {\it "Recapitulating, we may say that according to the general theory of relativity space is endowed with physical qualities; in this sense, therefore, there exists an ether. According to the general theory of relativity space without ether is unthinkable; for in such space there not only would be no propagation of light, but also no possibility of existence for standards of space and time (measuring-rods and clocks), nor therefore any space-time intervals in the physical sense. But this ether may not be thought of as endowed with the quality characteristic of ponderable media, as consisting of parts which may be tracked through time. The idea of motion may not be applied to it."} A different notion of the physical vacuum emerged with quantum mechanics and Dirac's electron-hole theory \cite{Dirac} leading to the discovery of the positron \cite{Anderson}. 

Quantum field theory confirmed the role of vacuum and brought the concept of vacuum polarization. Later, spontaneous symmetry breaking \cite{Nambu} and the Higgs mechanism \cite{Higgs} strengthened the idea of a material physical vacuum, where fields can condense. Similar to the ground state of condensed matter physics, the vacuum of particle physics is defined as the lowest-energy state of matter. Its excitations are assumed to be described by the standard particles of quantum field theory. But the validity of this approach at very high energy has not really been proven and requires experimental verification. 

If the vacuum is somehow the "ground state" of matter, it must in principle contain the most essential information on its ultimate structure and dynamics. Therefore, studying experimentally the actual properties of vacuum at very short distance scales can be an important challenge for high-energy cosmic-ray physics \cite{Gonzalez-Mestres2010}. 

\subsection{Validity of fundamental principles}

In 1921, Einstein wrote \cite{Einstein1921} about the application of relativity to the constituents of matter: {\it "It is true that this proposed physical interpretation of geometry breaks down when applied immediately to spaces of sub-molecular order of magnitude. But nevertheless, even in questions as to the constitution of elementary particles, it retains part of its importance. For even when it is a question of describing the electrical elementary particles constituting matter, the attempt may still be made to ascribe physical importance to those ideas of fields which have been physically defined for the purpose of describing the geometrical behaviour of bodies which are large as compared with the molecule. Success alone can decide as to the justification of such an attempt, which postulates physical reality for the fundamental principles of Riemann's geometry outside of the domain of their physical definitions. It might possibly turn out that this extrapolation has no better warrant than the extrapolation of the idea of temperature to parts of a body of molecular order of magnitude. It appears less problematical to extend the ideas of practical geometry to spaces of cosmic order of magnitude."} It is an extraordinary fact that, nine decades later and with data on elementary particles down to almost seventeen orders of magnitude below the size of a hydrogen atom, no violation of the Lorentz symmetry has been established. The Greisen-Zatsepin-Kuzmin (GZK) cutoff \cite{GZK1,GZK2}, if confirmed, would imply the success of calculations involving a $\approx ~ 6.10^{10}$ boost for protons and a $\approx ~ 10^{9}$ boost for iron nuclei. 

Again, high-energy cosmic-ray experiments turn out to be the only way to check the validity of a fundamental principle of physics at extreme scales. Ultra-high energy cosmic rays (UHECR) detected by earth-based experiments like AUGER \cite{AUG}, HiRes \cite{Hires} and the Telescope Array \cite{TA}, or by satellite missions like EUSO \cite{EUSO}, will remain unique instruments to test possible Lorentz symmetry violation (LSV) generated at the Planck scale or at some other fundamental scale \cite{Gonzalez-Mestres2008,Gonzalez-Mestres2009a}.

Together with LSV, but also perhaps independently, other violations of commonly admitted principles may manifest themselves through cosmic-ray physics \cite{Gonzalez-Mestres2010,Gonzalez-Mestres2009b}. They can possibly concern quantum mechanics, energy and momentum conservation, effective space-time dimensions, the validity of lagrangian and Hamiltonian formalisms, standard cosmology... including, as previously stressed, a possible composite structure of conventional particles or the properties of our physical vacuum and particle propagation. Unexpected discoveries in these domains would strongly influence not only the future and the basic hypothesis of particle physics, but also the fundamentals of cosmology. Issues like dark matter, dark energy, inflation, the cosmological constant or the existence itself of the big bang, would have to be reconsidered \cite{gonSL1,Gonzalez-Mestres2010,Gonzalez-Mestres2009a,Gonzalez-Mestres2009b}. 

Globally, the systematic study and design of possible tests of fundamental principles by cosmic-ray physics has just begun. It will become a central research field in the future.

\section{Lorentz symmetry violation (LSV)}

A discussion of possible implications of AUGER and HiRes data for LSV patterns was presented at the previous CRIS conference \cite{Gonzalez-Mestres2008}. Since then, HiRes has published its final results \cite{HiRes-final} claimed to be "completely consistent with a light, mostly protonic composition for the UHECR spectrum", whereas the Pierre Auger Collaboration \cite{AUGER0910} states that primary cosmic rays "are likely to be dominated by heavy nuclei at higher energies". Obviously, further data and analyses are required to settle this crucial issue.

As analyzed in \cite{Gonzalez-Mestres2008}, bounds on LSV strongly depend on the composition of the highest-energy cosmic-ray spectrum. Assuming that the observed fall of the UHECR spectrum above $E ~ \simeq ~ 10^{19.5}$ eV is due to the GZK cutoff, the bounds will be much more stringent if particles in this energy region are protons than in the case of heavy nuclei. Furthermore, as the AUGER Collaboration emphasizes \cite{AugerData}, the GZK cutoff is not the only possible conventional explanation of the UHECR flux suppression. Data could also reflect a maximum energy reachable at the sources. 

If the fall of the spectrum is due to a limitation of the existing sources, bounds on LSV can possibly be obtained taking into account the implications of the suppression of synchrotron radiation predicted by LSV models \cite{Gonzalez-Mestres2000b} and potentially allowing protons and nuclei to be accelerated to higher energies. Thus, a new branch of astrophysical LSV tests would be opened. Again, more data and analyses are required to check the usefulness and feasibility of such an approach.  

\subsection{LSV patterns}

We are interested in modifications of relativity that preserve Lorentz symmetry as a low-momentum limit, in a way compatible with existing bounds on LSV at low energies \cite{Lamoreaux}. 

To be able to produce observable effects in the UHECR region, models of deformed relativistic kinematics (DRK) must incorporate \cite{Gonzalez-Mestres2000a,gonSL1,gonLSV1} a preferred reference frame (the vacuum rest frame, VRF). Otherwise, a transformation to the center-of-mass frame of the interaction or to the rest frame of the single object under study would eliminate the effect. Models based on Finsler algebras or similar structures, where the laws of Physics are independent of the inertial frame considered \cite{Kir,SDSR}, could not explain phenomena like a possible absence of the GZK cutoff or a stability (unstability) of unstable (stable) particles at very high energy \cite{Gonzalez-Mestres2002,gonLSV2}. We call weak doubly special relativity (WDSR) the approach based on DRK and the existence of the VRF, contrary to standard doubly special relativity \cite{SDSR} that we call strong (SDSR). In WDSR, a particle with energy $\approx ~10^{20}$ eV in the VRF is not the same physical object as a similar particle at rest in this frame. Therefore, quark and gluon deconfinement in vacuum may occur above some energy threshold.

The Earth is usually assumed to move slowly with respect to the VRF. In the VRF with WDSR, a DRK can be formulated as follows:
\begin{equation}
E~=~~(2\pi )^{-1}~h~c~a^{-1}~e~(k~a)
\end{equation}
where $E$ is the particle energy, $a$ the fundamental length (Planck or another scale), $h$ the Planck constant, $c$ the speed of light, $k$ the wave vector, and $e~(k~a)$ a function incorporating the deformation of the kinematics. For $k~a~\ll ~1$, we get:
\begin{equation}
\begin{split}
e~(k~a) ~ \simeq ~ [(k~a)^2~-~\alpha ~(k~a)^{2+n}~ \\
+~(2\pi ~a)^2~h^{-2}~m^2~c^2]^{1/2}
\end{split}
\end{equation}
$\alpha $ being a model-dependent constant, {\it m} the mass and $n$ a positive exponent, integer in most cases. For momentum $p~\gg ~mc$ :
\begin{equation}
E ~ \simeq ~ p~c~+~m^2~c^3~(2~p)^{-1}~ \\
-~p~c~\alpha ~(k~a)^n/2~~~~~
\end{equation}

The deformation term $\Delta ~E~ \simeq ~-~p~c~\alpha ~(k~a)^n/2$ in (3) implies a LSV in the ratio $E~p^{-1}$ varying like $\Gamma ~(k)~\simeq ~\Gamma _0~k^n$ where $\Gamma _0~ ~=~-~\alpha ~c~a^n/2$. In terms of the fundamental energy scale $E_a ~ =~ h ~ c ~(2 ~\pi ~a)^{-1}$, equation (3) becomes:
\begin{equation}
\begin{split}
E ~ \simeq ~ p~c~+~m^2~c^3~(2~p)^{-1}~ \\
-~p~c~\alpha ~(p~ c ~E_a^{-1})^n/2~~~~~
\end{split}
\end{equation}
and $\Delta ~E~\simeq ~-~p~c~\alpha ~(p~ ~c ~E_a^{-1})^n/2$.
If $c$ is a universal parameter for all particles, the DRK defined by (1) - (4) preserves Lorentz symmetry in the limit $k~\rightarrow ~0$. $\alpha $ is usually taken to be positive and depends on the object considered \cite{Gonzalez-Mestres2008,gonLSV1}. For large composite structures of mass $M$, $\alpha $ would be proportional to $\simeq ~ M^{-n}$. Although we initially assumed for phenomenological purposes \cite{gonLSV1} the value of $\alpha $ to be basically the same for the all the standard "elementary" particles as well as for protons and neutrons, this hypothesis has been modified at a later stage \cite{Gonzalez-Mestres2008,gonLSV2}. In particular, the composite character of the proton must be fully taken into account to interpret current data. Nuclei have always been dealt with as composite objects with naturally smaller $\alpha $'s \cite{Gonzalez-Mestres2008,gonLSV1}.   

Kinematical balances and other basic properties of particle interactions are drastically modified when the deformation term becomes larger than the mass term $m^2~c^3~(2~p)^{-1}$, i.e. above the energy scale $E_{trans}~$ $\approx ~(\alpha ^{-1} ~ m^2 ~ c^4 ~ E_a^n)^{1/(2+n)}$. The internal structure of the particle can undergo a transition in this energy region \cite{Gonzalez-Mestres2008,Gonzalez-Mestres2002,Gonzalez-Mestres2004}. 

\subsection{QDRK}

Quadratically deformed relativistic kinematics (QDRK) corresponds to $n$ = 2 in (1) - (4). This seems to be the best suited choice for phenomenology \cite{Gonzalez-Mestres2000a,gonLSV1}, but it also naturally corresponds to composite pictures of the vacuum and of standard particles inspired by the solid-state Bravais lattice \cite{Gonzalez-Mestres2000a,gonSL1,gonLSV1} or by wave refraction with a Cauchy law \cite{Gonzalez-Mestres2010}. QDRK can naturally lead to the suppression of the GZK cutoff and to the stability of unstable particles at very low energy \cite{Gonzalez-Mestres1997}. Then, a similar mechanism can also suppress synchrotron radiation in particle acceleration to the same energies by astrophysical sources \cite{Gonzalez-Mestres2000b}. 

The choice $n$ = 1 (linearly deformed relativistic kinematics, LDRK) was discarded \cite{Gonzalez-Mestres2000a} in our phenomenological proposals for UHECR phenomenology, as it would lead to too strong effects at lower energies. It can be partially present in hybrid models with energy thresholds \cite{gonLSV2}. 

As pointed out in our CRIS 2008 talk \cite{Gonzalez-Mestres2008}, even assuming that the fall of the UHECR spectrum is due to the GZK cutoff, present data would not by themselves exclude a QDRK pattern with $\alpha ~\approx $ 0.1 or 1 for quarks and gluons corresponding to strong LSV at the Planck scale. For comparison, following an analogy with the Bravais lattice calculations for phonons \cite{Gonzalez-Mestres1997} would lead to $\alpha ~\approx $ 1/12 if $a$ is the equivalent of a lattice spacing. 

The GZK cutoff for a proton component of the UHECR spectrum would, if demonstrated, imply for $\alpha $ (proton) an upper bound $\approx ~10^{-6}$ if $a$ is the Planck length \cite{gonLSV1}. For quarks and gluons, this bound should probably be multiplied by $\approx ~N^2$, where $N$ is the number of effective constituents of the incoming protons. A $10^{20}$ eV iron nucleus would basically amount to a set of nucleons with energies $\simeq ~2.10^{18}$ eV. At these energies, the nucleon mass terms still dominate over the QDRK deformations for $\alpha $ (nucleon) $<$ 1 and $a$ = Planck length. Furthermore, the validity of present algorithms to estimate UHECR energy is not really established. It therefore seems necessary : i) to clearly identify a UHECR component lighter than iron; ii) to better understand UHECR interaction with the atmosphere, as well as the internal structure of UHECR nucleons; iii) to further explore and study UHECR sources and acceleration.

\section{Preons and superbradyons}

String models are often presented as the ultimate formulation of elementary particle physics. Some of them have been used to study possible deviations from standard Lorentz symmetry and astrophysical tests of these deviations \cite{SDSR,Mavromatos}. However, the complexity and structure of strings suggest the existence of an underlying composite dynamics \cite{gonSL3}. The string picture originated initially from the dual resonance models of hadronic physics \cite{DRM1}, and was then interpreted \cite{DRM2} in terms of "fishnet" Feynman diagrams involving quark and gluon lines. Current string patterns can be associated to possible superbradyonic "fishnet" diagrams \cite{Gonzalez-Mestres2009b,gonLSV2,Gonzalez-Mestres1997}. Superluminal constituents can directly replace strings at the Planck scale, or lead to an alternative theory.

In his December 1979 Nobel lecture, discussing the {\it "quest for elementarity"} and the preon model \cite{preons}, Abdus Salam emphasized that : {\it "quarks carry at least three charges (colour, flavour and a family number)"}. He suggested to {\it "entertain the notions of quarks (and possibly of leptons) as being composites of some more basic entities"} carrying each {\it "one basic charge"}. Subsequent developments led to more involved scenarios, but they did not raise the question of the validity of the fundamental principles of standard particle theories (special relativity, quantum mechanics...) for the new constituents. This was done for the first time in our papers since 1995 \cite{gonSL1,gonLSV1}, where the superbradyon hypothesis implied a radical change in the space-time structure felt by the new (non-tachyonic) superluminal particles, possible ultimate constituents or produced by a deeper composite structure. Superbradyons would have positive mass and energy, and a critical speed in vacuum $c_s ~\gg ~c$. With the suggestion of a superbradyonic sector of matter, it was stressed that: i) its interaction with ordinary matter would break standard Lorentz invariance ; ii) to be consistent with low-energy experiments, such a mixing would have to be a high-energy phenomenon. 

This 1995 scenario \cite{gonSL1} led to the DRK approach developed in 1997 \cite{gonLSV1}. Possible violations of standard quantum mechanics were not discarded and have been considered recently \cite{Gonzalez-Mestres2009b}.  
  
Assuming a kinematics of the Lorentz type with $c_s$ playing the role of the critical speed, the energy $E_s$ and momentum $p_s$ of a free superbradyon in the VRF would be given by \cite{gonSL1}: 
\begin{eqnarray}
E_s~=~c_s~(p_s^2~+~m_s^2 ~c_s^2)^{1/2} \\
p_s~=~m_s~v_s~(1 ~-~v_s^2~c_s^{-2})^{-1/2}
\end{eqnarray}
where $m_s$ is the superbradyon inertial mass and $v_s$ its speed. Actually, free superbradyons may undergo refraction in the physical vacuum of our Universe (like photons in condensed matter) or exist in it only as quasiparticles and other forms of excitations, or be confined, or be able to quit and enter this Universe \cite{Gonzalez-Mestres2010}. Then, the kinematics and critical speed of superbradyons in our vacuum would not be the same as in an "absolute" vacuum, assuming the latter can exist. But we shall not consider these complications here. 

Superbradyons can play an important role in cosmology \cite{gonSL1,Gonzalez-Mestres2008,Gonzalez-Mestres2009a,Gonzalez-Mestres2009b,gonSL2} and be a source of conventional UHECR through spontaneous decays ("Cherenkov" radiation in vacuum) \cite{gonSL2}. A superbradyonic era can even replace the standard Big Bang. When traveling at $v_s~>~c$, superbradyons obeying equations (5) - (6) would spontaneously emit standard particles until their speed becomes $\simeq ~ c$. They can form a cosmological sea and be candidates to dark matter and dark energy \cite{gonSL1,Gonzalez-Mestres2008,Gonzalez-Mestres2009a,Gonzalez-Mestres2009b,gonSL2}. Annihilation and decays of superbradyonic dark matter have been suggested \cite{Gonzalez-Mestres2009b} to explain cosmic positron abundance \cite{PAMELA}. Taking $v_s ~ \sim ~ c$ and $c_s ~ \sim ~10^6$ $c$ (similar to the ratio between $c$ and phonon speed), a superbradyon with $E$ in the TeV range would have a mass $\sim ~ 1$ eV $c^{-2}$, momentum $\sim ~ 1$ eV $c^{-1}$ and kinetic energy $\sim ~ 1$ eV. Such superbradyons would be very hard to detect, not only because of their expected very weak interaction with conventional matter but also because of the small available energy. Possible superbradyonic mixings in standard particles at LHC energies deserve further study \cite{gonSL3}. 

The situation would be different if the observed positron flux were due to "Cherenkov" emission by superbradyons with kinetic energy $~\sim ~1$ TeV, $v_s$ slightly above $c$ and $c_s ~ \sim ~ 10^6$ $c$. Then, the superbradyon rest energy would be $ m_s ~c_s^2 ~\sim ~10^{24}$ eV and some spectacular decays could perhaps be observed in UHECR experiments.

\section{Other tests of basic principles}

The proposed approach based on QDRK and the superbradyon hypothesis is not a purely phenomenological one. It incorporates a coherent set of basic hypotheses implying a composite character of conventional particles and a deformation of the relativistic kinematics consistent with composite pictures (phonon-like or refraction-like dispersion relations). It therefore contains the embryo of a new fundamental theory, to be made more precise as information from UHECR experiments and from other sources will help to clarify the situation. Superbradyonic physics can be substantially different from standard particle theory, and superbradyons are just an illustrative example of possible new physics beyond Planck scale. Therefore, all conventional fundamental principles require further experimental verification.    

QDRK can suppress the GZK cutoff, but it can also generate mechanisms faking this cutoff and based, for instance, on spontaneous UHECR decays due to differences in the value of $\alpha $ between standard particles \cite{Gonzalez-Mestres2008,Gonzalez-Mestres2009a,Gonzalez-Mestres2009b,gonLSV1}. Similar effects, combined with LSV or independent of it, can result from other violations of fundamental principles such as quantum mechanics ($h$ has a comparatively large uncertainty) or energy-momentum conservation \cite{Gonzalez-Mestres2009b}, or from unexpected vacuum properties (local fluctuations, energy capture or release) \cite{Gonzalez-Mestres2010}. Deformed quantum commutation relations can lead to intrinsic uncertainties in energy and momentum for UHECR \cite{Gonzalez-Mestres2009b}. 

Testing experimentally the actual structure of space-time is a natural question in most theoretical approaches to particle physics, not only about the validity of Lorentz symmetry but more generally. A possibility considered in \cite{gonSL2,gonSL4} was to replace the standard four-dimensional space-time by a SU(2) spinorial one, so that spin-1/2 particles would be representations of the actual group of space-time transformations. Extracting from a spinor $\xi $ the scalar $\mid \xi \mid ^2$ $=$ $\xi ^\dagger \xi $ where the dagger stands for hermitic conjugate, a positive cosmic time $t~=~\mid \xi \mid$ can be defined which leads in particular to a naturally expanding Universe. 

Another unconventional space-time pattern has been recently suggested by Anchordoqui and other authors \cite{Landsberg}, where the number of effective space dimensions decreases with the energy scale through scale thresholds. Possible thresholds in LSV were also considered in \cite{gonLSV2}. As already foreseen for LSV and DRK \cite{Gonzalez-Mestres2000a,gonLSV1,gonLSV2}, this new LSV approach may cure ultra-violet divergencies in field theories. Anchordoqui et al. also suggest that the pattern presented in \cite{Landsberg} may explain elongated jets in cosmic-ray data possibly observed by Pamir \cite{Pamir} and other experiments. 

As shown in \cite{Gonzalez-Mestres2010}, missing transverse energy in cosmic-ray interaction jets above some energy scale ($\sim ~ 10^{16}$ eV ?) can actually be a consequence of the production of superluminal objects (waves, particles...) involving a small portion (provided by the target) of the total energy and a negligible fraction of momentum. Assume for simplicity that a UHECR of mass {\it m}, energy $E$ and momentum $p$ hits an atmospheric target of mass {\it M} at rest, and that the final state is made of two particles of mass $m'$ and longitudinal momentum (in the direction of the incoming cosmic ray) $p/2$. The total energy is $E~+~M~c^2$, and $E$ will be mainly spent to fulfil the requirement of momentum conservation in the longitudinal direction. We get for each of the two produced particles:
\begin{equation}
{p_T}^2~\simeq ~M~c~p/4 
\end{equation}   
where $p_T$ is transverse momentum, corresponding to a  transverse energy $E_T~\simeq ~M~c^2/2$ provided by the target mass term. The available transverse energy for these secondaries becomes smaller in the presence of a simultaneous emission of exotic objects (particles, waves...) with an overall energy (captured by vacuum?) $\Delta E_{vac}$ comparable to that of the target and longitudinal momentum $\Delta p_{vac} ~<<~\Delta E_{vac} ~c^{-1}$. The transverse energy of the above secondaries is then $E_T~\simeq ~(M~c^2~-~\Delta E_{vac})/2$, with:
\begin{equation}
{p_T}^2~\simeq ~(M~c~-~\Delta E_{vac} ~ c^{-1})~p/4   
\end{equation}   
Superbradyonic kinematics would forbid a significant momentum for the exotics. $v_s ~ \sim ~ c$ and $c_s ~ \sim ~10^6$ $c$ yield $m_s~\sim ~ 10^{-3}$ eV $c^{-2}$, $p_s ~\sim ~ 10^{-3}$ eV $c^{-1}$ and kinetic energy $\sim ~ 10^{-3}$ eV \cite{Gonzalez-Mestres2010}. Polarization effects inside vacuum and secondaries can play a role in subsequent planar jet alignment.

\section{Conclusion} 

When suggesting to test relativity through LSV patterns with a VRF and UHECR experiments, it was stressed more generally \cite{Gonzalez-Mestres2000a} that high-energy cosmic-ray physics provides a powerful microscope directly focused on the fundamental length (Planck?) scale. Such a statement applies in fact to all basic principles of Physics. The efficiency of this unprecedented tool will depend on the amount and quality of UHECR data, as well as on our understanding of these data and of other physical informations. This will necessarily require a long-term effort before trying to build a realistic new theory of matter and space.

\section{Post Scriptum after publication in Nuclear Physics Proceedings}

\subsection{SL(2,C), SU(2), SO(4), causality, half-integer spin and Cosmology}  
\vskip 2mm
To define local space coordinates in the above spinorial approach to space-time \cite{gonSL4}, one can consider a spinor $\xi _0$ (the observer position) on the $\mid \xi \mid $ = $t_0$ hypersphere. Writing, for a point $\xi $ of the same spatial hypersphere :
\begin{equation}
\xi ~=~ U ~ \xi _0
\end{equation}
where $U$ is a cosmic SU(2) transformation :

\begin{equation}
U~=~exp~(i/2~~t_0 ^{-1}~{\vec \sigma }.{\vec {\mathbf x}})~
\equiv U({\vec {\mathbf x}}) 
\end{equation}
and ${\vec \sigma }$ the vector formed by the Pauli matrices, the vector ${\vec {\mathbf x}}$, with $0~\leq x$ (modulus of $\vec {\mathbf x}$)  $\leq$  $2\pi t_0$, can be interpreted as the cosmic spatial position vector of $\xi $ with respect to $\xi _0$ at constant time $t_0$. 

A $2\pi $ SU(2) rotation in the spinorial space-time ($x~=~2\pi t_0~,~U~=~-1$), or the equivalent change of coordinates, changes the signs of $\xi _0$ and $\xi $ simultaneously. It therefore turns $\xi _0$ and $\xi $ into their cosmic antipodals but leaves ${\vec {\mathbf x}}$ invariant. Thus, the relevant variation domains for $x$ before reaching again the identity map are $0~\leq x ~\leq$  $2\pi t_0$ in the case of space coordinates defined at constant $t_0$, and $0~\leq x ~\leq$  $4\pi t_0$ for space-time spinors.

Conventional ("euclidean") local space coordinates are obtained for $x ~\ll ~t_0$ where the effects of the noncommutativity of the Pauli matrices can be neglected. 

Thus, space translations are just SU(2) transformations acting on the space-time spinors whereas standard space rotations act on the SU(2) transformations themselves. 
\vskip 2mm
{\it 6.2.a A simple cosmological check}

Assuming that the $\mid \xi \mid $ = $t_0$ hypersphere corresponds to the three-dimensional space when the age of our Universe is $t_0$ naturally leads to the so-called Hubble's law, actually first formulated by Georges Lema\^itre \cite{Lemaitre1927}, relating distances and relative velocities.

More precisely, assuming that the ratio between the physical space units and time units remains constant with cosmic time, and that the cosmic time units correspond to the time units of observations, the value of the Hubble (Lema\^itre) constant would be exactly equal to the inverse of the age of the Universe $t_0$. The expansion speed would then be constant with time (vanishing deceleration parameter). Similarly, redshifts are automatically generated by the expansion of the Universe in the spinorial space-time.

Such a structure is obtained on very basic and general geometric grounds, without introducing any specific cosmology and with even no space units or Minkowskian metric. At that stage, the four components of the two complex spinorial coordinates are measured with a single (time) scale, and the only specific space coordinates are SU(2)-angular and comoving.

Actually, the large-scale structure of our Universe appears compatible with a SO(4) symmetry where, instead of being imaginary, the cosmic time would correspond to the modulus of a four-vector. SO(4) would unify, in this context, translations in the curved space with the standard space rotations. However, as shown above and further discussed below, translations and space rotations should actually be viewed as different representations of a single SU(2) group. 

A transition from such a cosmic SU(2) or SO(4) inherited from the initial singularity $\xi $ = 0 to a "dynamical" symmetry including SL(2,C) at the nowadays local space-time scales is obviously possible, but it would require further physical and geometrical input.

Thus, we have produced in a very simple way the pattern of a potentially ever-expanding universe with positive (3-spherical) curvature that may be basically compatible with experimental data and deserves being further explored. Contrary to cosmologies where a fourth Euclidean coordinate is related to an imaginary time, a cosmic time is defined here as the modulus of a SU(2) spinor or that of the associated real four-vector.

As we have not yet introduced in the model matter and its interactions, or even the precise connection between space units and angular commoving coordinates at a given time, we shall not try to discuss here topics like the possible acceleration of the expansion of the Universe, dark matter or dark energy. We defer this to a later work. 

Light waves for a zero-mass photon with a critical speed $c$ can be introduced without any explicit reference to Lorentz symmetry, just by defining the energy as the product of the speed of light by the inverse of the wavelength. The need for standard relativity arises when considering massive conventional particles of light emission by moving sources.
\vskip 2mm
{\it 6.2.b Other space-time transformations}

Using the above conventions, a standard space rotation around $\xi _0 $ is defined by a SU(2) element $U({\vec {\mathbf y}})$ turning any $U({\vec {\mathbf x}})$ into $U({\vec {\mathbf y}}) ~ U({\vec {\mathbf x}}) ~ U({\vec {\mathbf y}})^\dagger $. The vector $\vec {\mathbf y}$ provides the rotation axis and angle. As already stated, a $2\pi $ rotation ($x ~=~ 2\pi t_0$ ) leaves $\vec {\mathbf x}$ invariant.

Translations are therefore "cosmic" rotations of the spinors around the "initial" point of the Universe $\xi ~=~ 0$, whereas standard space rotations are SU(2) transformations acting on the set of local space translations from a fixed point $\xi _0$. 

But another kind of local position can also be defined if the requirement of constant cosmic time does not apply to the path between $\xi _0$ and $\xi $ : the "direct" spinorial separation $\Delta \xi ~= ~\xi ~-~ \xi _0$, which changes sign under a $2\pi $ SU(2) rotation in the spinorial space-time. A spinorial SU(2) rotation around $\xi _0$ is then constructed by transforming the point $\xi $ into $\xi '$ = $\xi _0$ + $U({\vec {\mathbf y}})~\Delta \xi $ . Obviously, the spinorial position behaves differently from the standard space position under a $2\pi $ rotation. As the spinorial position can in principle be used to build wave functions similar to the standard vector position, we should expect half-integer angular momenta to be realized in Nature. 

It must however be noticed that the fact that such a spinorial position is not defined through a constant-time link may have important negative consequences potentially limiting its role in standard physics, even if the position $\Delta \xi $ clearly exists from a mathematical point of view and transforms like a spinor under SU(2) rotations. Strictly speaking, a "straight line" between $\xi _0$ and $\xi $ in the spinorial space-time would not be really compatible with standard causality in our classical macroscopic cosmic picture, due to the very small effect from the Universe spatial curvature. Contrary to the standard space position, a SU(2) spinorial rotation around $\xi _0$ does not necessarily tranform $\xi $ into an element of the $\mid \xi '\mid $ = $t_0$ hypersphere. The image of $\Delta \xi $ under SU(2) transformations is actually a S$^3$ hypersphere similar to the spinorial Universe itself but centered around $\xi _0$.

More generally, any effective space curvature at constant time, even local, can potentially lead to the same kind of phenomenon and obstruct the "direct" spinorial link. 

This circumstance may explain the absence in Nature of half-integer angular momenta other than those associated to the spin of "elementary" fermions. One can conceive that the "direct" spinorial position is relevant only at extremely small distance scales, where the internal structure of conventional "elementary" particles is formed (Planck or beyond ?) and standard causality or similar constraints, including the arrow of time, do no longer hold. The constant-time space coordinates considered above cannot lead to a half-integer spin.

Thus, the spin 1/2 could in principle be naturally generated as angular momentum of the matter in motion forming our standard particles at distance scales where causality and other standard requirements do not apply and the description of kinematics can be different ("inside" an electron or a quark). 

The existence of half-integer spin may then be interpreted as a possible evidence for causality-violating physics (or some equivalent change in the effective space-time structure felt by matter) at ultimate length and space scales, if one assumes that such a spin is indeed some sort of physical "orbital" angular momentum corresponding to the fundamental representation of SU(2). Quantum uncertainties or pre-quantum physics and similar phenomena can naturally play a role at these very small scales, even if half-integer spin is not generated in the conventional quantum mechanics using standard space coordinates and rotations based on our macroscopic perception of space and time.

Similar considerations also suggest that the region between $\xi $ = 0 and, roughly, the $\xi $ = $t_{Planck}$ (Planck time) hypersphere can naturally correspond to a pre-Big Bang scenario taking $\xi $ = 0 as the initial singularity or starting from a (superbradyonic ?) time scale much smaller than the Planck time. Contrary to existing prejudices, tests of pre-Big Bang physics can in principle be possible with WMAP and Planck data, but also through UHECR experiments where extrapolations from physics beyond Planck scale may be detectable.  

In our SU(2) Universe, SL(2,C) transformations introduce basically a change of relative scale between the two components of the cosmic spinor taking the SL(2,C) transformation to be diagonal. Similar transformations can be considered if the origin is the point $\xi _0$ instead of $\xi $ = 0 .
\vskip 2mm
{\it 6.2.c A privileged space direction ?}

A specific property of the spinorial space-time considered here is that, to each point $\xi $ , a (complex) one-dimensional spinorial subspace can be associated such that for any point $\xi '$ of this subspace one has : 
\begin{equation}
\xi ^\dagger ~ \xi ' ~ = ~ \mid \xi '\mid  ~ \mid \xi \mid ~ exp ~(i \phi )
\end{equation}
where $exp ~(i \phi )$, with $\phi $ real, stands for a complex phase.

If $\mid \xi '\mid  ~ = ~ \mid \xi \mid $ so that $\xi $ and $\xi '$ belong to the same constant-time hypersphere, the complex phase $exp ~(i \phi )$ is associated to the SU(2) matrix transforming $\xi $ into $\xi '$. This transformation, $U~=~exp~(i/2~~t ^{-1}~{\vec \sigma }.{\vec {\mathbf z}})$ where $t$ is the cosmic time $t ~= ~ \mid \xi \mid $ and ${\vec {\mathbf z}}$ a real space vector, is generated by a sigma-like matrix ${\vec \sigma }.{\vec {\mathbf z}}~ \mid z \mid ^{-1}$ associated to a unique space direction ${\vec {\mathbf z}}~ \mid z \mid ^{-1}$ on the constant time hypersphere. 

$\xi $ and $\xi '$ are both eigenspinors of ${\vec \sigma }.{\vec {\mathbf z}}$ . For each point $\xi $ of the spinorial space-time, other than $\xi $ = 0, there exists a unique space direction for which $\xi $ is an eigenspinor of the associated sigma-like matrix. Exponentiating this matrix with an imaginary coefficient generates the directions of the relevant (complex one-dimensional) spinorial subspace associated to $\xi $ .

With $\mid \xi '\mid  ~ = ~ \mid \xi \mid $ and a positive phase $\phi $, one actually has :
\begin{equation}
\xi ' ~ = ~ exp ~(i/2~~t ^{-1}~ \mid z \mid )~ \xi  
\end{equation}
and similarly, with $ - \mid z \mid $ instead of $\mid z \mid $, if $\phi $ is negative. 

The set of points of the spinorial space-time thus generated obviously corresponds to a (spinorial) circle of radius $\mid \xi '\mid ~=~ t$ ($t$ = cosmic time) on the constant-time hypersphere, including the point $\xi $ itself and its SU(2) antipodal - $\xi $ .

Thus, "looking at" the initial point of our Universe $\xi $ = 0 from a point $\xi $ of the present time spatial hypersphere naturally leads, in the spinorial coordinates considered here, to the definition of a privileged space direction on the space hypersphere itself.

The direct memory of the geometry leading to such a privileged space direction is basically lost if standard space coordinates on the constant-time hypersphere are used and standard matter is dealt with without incorporating its deepest structure as well as the most primordial origin of the Universe. However, several possible tracks from this spinorial effect in Cosmology and Particle Physics can still be considered. 

In particular :

- The internal structure of standard spin-1/2 particles, as well as their interaction properties at very small distance scales, may contain the expression of a similar phenomenon.

- Signatures from a pre-Big Bang era can yield relevant information on this privileged space direction and on effects of the same origin through WMAP, Planck and other experiments.

- Similarly, ultra-high energy cosmic rays may be sensitive to both cosmological and "beyond Planck" phenomena containing effects related to the privileged space direction.

Further work on this subject is clearly required.

\subsection{From SL(2,C) to SL(N,C) ?}
\vskip 2mm 

To the above SU(2) symmetry, a natural SL(2,C) is tacitly associated, potentially including Lorentz transformations in the usual way except possibly for the definition of the time variable. This point will be further discussed elsewhere. Then, $\mid \xi \mid ^2$ $=$ $\xi ^\dagger \xi $ is no longer a scalar, and cosmic time can be defined only in a preferred reference frame. Lorentz symmetry may be preserved in the limit where Physics does not depend on the choice of this frame and on the above definition of the cosmic time.  

As in this approach space translations and rotations are not fundamentally different transformations, the question whether some suitable extension of SL(2,C) can meet less stringent no-go constraints than those stated some decades ago \cite{nogo} deserves being addressed. 

We suggest, in particular, to consider in this kind of picture some broken version of SL(N,C) as the possible global symmetry of the Physics of standard particles. 

In a simple picture, SL(N,C) would act on the two complex dimensions associated to the usual space-time, plus N-2 complex spinorial coordinates related to extra dimensions and to internal symmetries. N-spinors with nonvanishing coordinates only in the N-2 "non space-time" dimensions would be scalars under the standard space-time SL(2,C). Transformations inside the new N-2 dimensions would correspond to internal symmetries.

Similarly, transformations between the standard SL(2,C) space-time spinors and the spinors oriented in the new N-2 dimensions would be of supersymmetric nature.

A more detailed discussion of SL(N,C) potentialities will be presented in a forthcoming paper.

\subsection{SO(N) and other groups}
\vskip 2mm 
Similar to the SO(4) discussed above, other SO(N) symmetries can be considered as extensions of the pattern presented.

The geometric universe considered using SU(2) and possibly SO(4) can also be identified to the subset of null (zero-norm) vectors in a SO(4,1) pattern where the fifth dimension would be the cosmic time, and the metric :
\begin{equation}
X^2~=~ \xi ^\dagger \xi ~- ~t^2
\end{equation}

\noindent where $X$ is the 5-vector formed by the four real components of the space-time spinor, and $t$ the cosmic time defined as the modulus of this space-time spinor.

Local versions of the same symmetry centered around a SU(2) spinorial point $\xi _0$ different from $\xi $ = 0 also deserve consideration, even if in this case the direct physical interpretation the $\mid \xi ~- ~ \xi _0\mid $ can appear much less obvious. The question of whether $\mid \xi ~- ~ \xi _0\mid $ can be interpreted at very short distances as the "internal time" of an "elementary" particle deserves being addressed.

In the SO(4,1) associated to the metric defined by (11), invariances of the Lorentz type naturally appear writing for instance :
\begin{equation}
x_1'~=~ \gamma ~x_1 ~- (\gamma ^2~- ~1)^{1/2} ~t
\end{equation}
\begin{equation}
t'~=~ \gamma ~t ~- (\gamma ^2 ~-~1)^{1/2} ~x _1 
\end{equation}
\noindent where $x_1$ is one of the four real coordinates extracted from the spinor $\xi $ and $x_1'$, $t'$ the transformed values of $x_1$, $t$. $\gamma$ is taken to be real and positive, with $\gamma ~>~1$. No reference to a critical speed, nor even to space units, is required in the formulation of such trivial Lorentz-like transformations.

\subsection{What does the concept of symmetry become at ultra-high energy ?} 
\vskip 2mm 
Is the Planck scale the ultimate limit where all symmetries are unified and become exact ? This standard picture is often considered as natural, and even the deformations of relativity by quantum gravity are turned into a new (deformed) symmetry where all inertial frames are equivalent. The possibility that new Physics can exist beyond the Planck scale is often just neglected. 

However, in our Universe there is a natural reference frame provided by the cosmic microwave background radiation. Furthermore, it is by now impossible to exclude on experimental grounds a real breaking of Lorentz symmetry as considered in this paper and in the references quoted above. There is also by now no proof of the validity of currently fashionable grand-unification patterns, and no supersymmetric particle has yet been found in spite of many expectations.

Furthermore, we have just suggested a pre-cosmological geometry that reasonably incorporates several main features of the large-scale properties of our Universe without introducing standard matter and its interactions. It may therefore happen that standard particle physics is not at the origin of the deepest structure of the Universe at very large scales, and that it simultaneously fails to account for the properties of matter at the really fundamental level.

An alternative view to the standard picture of particle symmetries has been suggested in the Post Scripta to \cite{Gonzalez-Mestres2009a} and \cite{Gonzalez-Mestres2009b}. If the standard particles have a composite origin and the vacuum is not made of conventional matter, it may happen that the usual gauge principle has no real physical meaning at the ultimate fundamental scale and that the gauge bosons are generated by the fundamental matter only in specific situations involving the physical presence of conventional particles as vacuum excitations. Similarly, the Higgs boson would not need to be statically materialized in vacuum as a permanent condensate in the absence of surrounding standard particles. 

In such a scenario, as energy increases beyond some scale that starts being sensitive to the internal structure of the standard "elementary" particles, differences between kinds of particles can become more and more detectable. The apparent symmetry may then tend to disappear or remain only as a low-energy approximation. Similarly, as considered above, the vacuum may start developing unexpected properties and other fundamental principles can possibly fail.

In other words, the standard concept of symmetry would be a mathematical construction to account for the intrinsic limitations of our low-energy view of matter. Similarly, for quantum field theory. Particles whose internal differences at a smaller distance scale cannot be observed by low-energy physics are described as "identical" in the sense of symmetry (including the generation pattern), even if they are actually not when examined at a deeper level. 

Then, instead of considering {\it a priori} that symmetries will become more and more exact as the energy scale increases and masses can be neglected as compared to the energy scale, one can conceive that this behaviour of the laws of Physics will exist in some range of scales but will change above a transition energy scale. Above the transition energy, observations would then become sensitive to new features of the particle internal structure and to properties of the real fundamental matter, increasingly unveiling differences in structure among particles from the same symmetry multiplet. Just below this transition scale, standard symmetries will reach their maximal reliability and precision.

Such an energy dependence of the properties of particle symmetries would also help to escape strong no-go conditions on the content of the allowed symmetries of conventional particles. From this point of view, a SL(N,C) symmetry can be considered as a tool and as a framework where to examine representations, symmetry breaking patterns, dynamical concepts and other relevant properties of Physics on a large set of scales, including practical no-go constraints.


\begin{thebibliography}{99}

\bibitem{Poincare1895} H. Poincar\'e, {\it L'\'eclairage \'electrique} Vol. 5, 5 (1895).

\bibitem{Poincare1902} H. Poincar\'e, "La science et l'hypoth\`ese", Flammarion, Paris 1902.
\bibitem{Gonzalez-Mestres2000a} L. Gonzalez-Mestres, ICRC 1999 Evening Workshop Session, arXiv:physics/0003080 and references therein.
\bibitem{Logunov} A.A. Logunov: "Henri Poincar\'e and Relativity Theory", arXiv:physics/0408077 and references therein ; R.P. Feynman, "The Character of Physical Law", BBC 1965.
\bibitem{Nambu} Y. Nambu, "Spontaneous symmetry breaking in particle physics: a case of cross fertilization", 2008 Nobel Lecture, and references therein. 
\bibitem{gonSL1} L. Gonzalez-Mestres, Moriond Workshop on "Dark Matter in Cosmology, Clocks and Tests of Fundamental Laws", Villars (Switzerland), January 1995, Ed. Fronti\`{e}res, arXiv:astro-ph/9505117 ; "Vacuum Structure, Lorentz Symmetry and Superluminal Particles", arXiv:physics/9704017 ; Workshop on "Observing Giant Cosmic Ray Air Showers for $> 10^{20} eV$ Particles from Space", Univ. of Maryland, November 1997,  {\it AIP Conf. Proc.} {\bf 433}, 418, arXiv:physics/9712049 and other papers by the same author.
\bibitem{Einstein1920} A. Einstein, "\"{A}ther und Relativit\"{a}tstheorie", Leyden May 5th, 1920. Translation in "Sidelights on relativity", Methuen, London 1922.
\bibitem{Dirac} P.A.M. Dirac, {\it Proc. R. Soc. Lond.} {\bf A126}, 360 (1930); {\it ibid.} {\bf A133}, 60 (1931). 
\bibitem{Anderson} C.D. Anderson, {\it Phys. Rev.} {\bf 43}, 491 (1933). 
\bibitem{Higgs} F. Englert and R. Brout, {\it Phys. Rev. Lett.} {\bf 13}, 321 (1964); P.W. Higgs, {\it Ibid.} {\bf 13}, 508 (1964); G. S. Guralnik, C. R. Hagen, and T. W. B. Kibble, {\it Ibid.} {\bf 13}, 585 (1964).
\bibitem{Gonzalez-Mestres2010} L. Gonzalez-Mestres, "Lorentz violation, vacuum, cosmic rays, superbradyons and Pamir data", arXiv:1009.1853
\bibitem{Einstein1921} A. Einstein, {\it Preus. Akad. der Wissench., Sitzungsberichte}, part {\bf I}, p. 123 (1921). Translation in "Sidelights on relativity", Methuen, London 1922. 
\bibitem{GZK1} K. Greisen, Phys.Rev.Lett. {\bf 16}, 148 (1966).
\bibitem{GZK2} G.T. Zatsepin and V.A. Kuzmin, JETP Lett. {\bf 4} , 78 (1966).
\bibitem{AUG} Pierre Auger Observatory, http://auger.org/ 
\bibitem{Hires} High Resolution Fly's Eye (HiRes), http://www.cosmic-ray.org/
\bibitem{TA} Telescope Array and TALE, http://www.telescopearray.org/
\bibitem{EUSO} Extreme Universe Space Observatory (EUSO), http://euso.riken.go.jp/
\bibitem{Gonzalez-Mestres2008} L. Gonzalez-Mestres, "Lorentz symmetry violation and the results of the AUGER experiment", arXiv:0802.2536 ;  CRIS (Cosmic Ray International Seminar), La Malfa (Italy), september 2008, {\it Nucl. Phys. B - Proc. Suppl.} {\bf 190}, 191, arXiv:0902.0994
\bibitem{Gonzalez-Mestres2009a} L. Gonzalez-Mestres, Invisible Universe International Conference, Paris, 29 June-July 2009, {\it AIP Conf.Proc.} {\bf 1241}, 1207, arXiv:0912.0725 and references therein.
\bibitem{Gonzalez-Mestres2009b} L. Gonzalez-Mestres, "Superbradyons and some possible dark matter signatures", arXiv:0905.4146 ; "Preon models, relativity, quantum mechanics and cosmology (I)", arXiv:0908.4070 ; and references therein.
\bibitem{HiRes-final} P. Sokolsky for the HiRes collaboration, these Proceedings and ISVHECRI 2010, Batavia, IL, USA, June - July 2010, arXiv:1010.2690
\bibitem{AUGER0910} The Pierre Auger Collaboration, "Update on the correlation of the highest energy cosmic rays with nearby extragalactic matter", arXiv:1009.1855
\bibitem{AugerData} The Pierre Auger Collaboration, Phys.Lett. {\bf B685}, 239 (2010), arXiv:1002.1975; S. Andringa for the Pierre Auger Collaboration, Rencontres de Moriond 2010 (Electroweak), arXiv:1005.3795 ; J.W. Cronin, Rencontres de Blois "Windows on the Universe", June 2009, arXiv:0911.4714 ; and references therein.
\bibitem{Gonzalez-Mestres2000b} L. Gonzalez-Mestres, International Symposium on High Energy Gamma-Ray Astronomy, Heidelberg, Germany, June 2000, {\it AIP Conf.Proc.} {\bf 558}, 874, arXiv:astro-ph/0011182
\bibitem{Lamoreaux} See, for instance, S.K. Lamoreaux, {\it Nature} {\bf 416}, 803 (2002) and references therein.
\bibitem{gonLSV1} L. Gonzalez-Mestres, 25th ICRC, Durban 1997, arXiv:physics/9705031 and pre-conference session, arXiv:physics/9706022; International Conference on Relativistic Physics and some of its Applications, Athens, June 1997, arXiv:physics/9706032 and arXiv:physics/9709006 ; EPS-HEP97 Conference, Jerusalem, August 1997, arXiv:nucl-th/9708028 ; Workshop on "Observing Giant Cosmic Ray Air Showers for $> 10^{20} eV$ Particles from Space", Univ. of Maryland, November 1997, {\it AIP Conf. Proc.} {\bf 433}, 148, arXiv:physics/9712047 and other papers by the same author.
\bibitem{Kir} D.A. Kirzhnits, and V.A. Chechin, V.A., ZhETF Pis. Red. {\bf 4} , 261 (1971); Yad. Fiz. {\bf 15} , 1051 (1972).
\bibitem{SDSR} See for instance : G. Amelino-Camelia, {\it Symmetry} {\bf 2}, 230 (2010), arXiv:1003.3942, and 2nd Mexican Meeting on Mathematical and Experimental Physics, Mexico City September 2004, {\it AIP Conf. Proc.} {\bf 758}, 30, gr-qc/0501053 ; J. Lukierski, SQS'03 Workshop, Dubna, Russia, July 2003, hep-th/0402117 ; J. Magueijo, {\it Phys.Rev.} {\bf D73}, 124020, (2006), gr-qc/0603073 ; L. Smolin, in "Approaches to Quantum Gravity", Ed. D. Oriti, Cambridge University Press, hep-th/0605052 , and "Could deformed special relativity naturally arise from the semiclassical limit of quantum gravity?", arXiv:0808.3765 ; J. Kowalski-Glikman, 22nd Max Born Symposium Wroclaw, Poland, September 2006, hep-th/0612280 ; and references therein.
\bibitem{Gonzalez-Mestres2002} L. Gonzalez-Mestres, "Deformed Lorentz Symmetry and High-Energy Astrophysics" (II) and (III), hep-th/0208064 and hep-th/0210141 , and references therein.
\bibitem{gonLSV2} L. Gonzalez-Mestres, updated material presented a the 35th COSPAR Scientific Assembly, Paris July 2004, hep-ph/0510361 ; I Symposium on European Strategy for Particle Physics, Orsay 2006, hep-ph/0601219
\bibitem{Gonzalez-Mestres1997} L. Gonzalez-Mestres, "Vacuum Structure, Lorentz Symmetry and Superluminal Particles", april 1997, arXiv:physics/9704017 and subsequent papers by the same author.
\bibitem{Gonzalez-Mestres2004} L. Gonzalez-Mestres, 28th ICRC, Tsukuba 2003, arXiv:hep-th/0407254 , arXiv:hep-ph/0407335 and arXiv:astro-ph/0407603
\bibitem{Mavromatos} See, for instance, N. Mavromatos, "String Quantum Gravity, Lorentz-Invariance Violation and Gamma-Ray Astronomy", arXiv:1010.5354 and references therein.
\bibitem{gonSL3} L. Gonzalez-Mestres, EPS-HEP97, Jerusalem August 1997, arXiv:physics/9708028
\bibitem{DRM1} See for instance "String Theory and Fundamental Interactions", {\it Lecture Notes in Physics} {\bf 737}, Ed. P. Di Vecchia and Jnan Maharana, Springer 2008. 
\bibitem{DRM2} H. B. Nielsen and P. Olesen, {\it Phys. Lett.} {\bf 32B}, 203 (1970); B. Sakita and M.A. Virasoro, {\it Phys. Rev. Lett.} {\bf 24}, 1146 (1970). 
\bibitem{preons} A. Salam, "Gauge Unification of Fundamental Forces", 1979 Nobel lecture, and references therein.
\bibitem{gonSL2} L. Gonzalez-Mestres, 4{th} TAUP Workshop, Toledo September 1995, {\it Nucl. Phys. Proc. Suppl.} {\bf 48}, 131, arXiv:astro-ph/9601090 ; "Superluminal Matter and High-Energy Cosmic Rays", arXiv:astro-ph/9606054 ; 28{th} ICHEP, Warsaw 1996, arXiv:hep-ph/9610474 and subsequent papers by the same author.
\bibitem{gonSL4} L. Gonzalez-Mestres, arXiv:astro-ph/9702026 "Space, Time and Superluminal Particles".
\bibitem{PAMELA} See, for instance, M Boezio et al. (PAMELA experiment), New J. Phys. {\bf 11}, 105023, http://iopscience.iop.org/1367-2630/11/10/105023 and references therein.
\bibitem{Landsberg} L. Anchordoqui et al., "Vanishing Dimensions and Planar Events at the LHC", arXiv:1003.5914, and references therein.
\bibitem{Pamir} I.P. Ivanenko et al., Stanford preprint SU-ITP-93-20, arXiv:hep-ph/9308218 ; V. Kopenkin et al., Phys.Rev. D {\bf 52}, 2766 (1995).
\bibitem{nogo} S. Coleman and J. Mandula, Phys. Rev. {\bf 159}, 1251 (1967) ; R. Haag, J. T. Lopuszanski and M. Sohnius, Nucl. Phys. {\bf B 88}, 257 (1975). 
\bibitem{Lemaitre1927} G. Lema{\^i}tre, "Un univers homog\`ene de masse constante et de rayon croissant, rendant compte de la vitesse radiale des n\'ebuleuses extra-galactiques", Ann. Soc. Sci. Brux. {\bf A 47}, 49 (1927), http://adsabs.harvard.edu/abs/1927ASSB...47...49L 

\end{thebibliography}
\end{document}